\documentclass[fullpage, 12pt]{article}
\title{Design-Aware Variance Reduction for Switchback Experiments: A Comparative Study}

\author{Sergei Pankratev \\ DoorDash, Inc.}
\date{}

\usepackage[a4paper, total={6in, 10in}]{geometry}
\usepackage{amsmath, amssymb}
\usepackage{booktabs}
\usepackage{graphicx}
\usepackage{float}
\usepackage[T1]{fontenc}
\usepackage{lmodern}
\usepackage{microtype}
\usepackage{threeparttable}
\usepackage{tabularx}
\usepackage{subcaption}
\usepackage[inline]{enumitem}
\usepackage{xcolor}
\usepackage{natbib}
\usepackage{hyperref}

\hypersetup{
colorlinks, linkcolor={red!
50!
black}, citecolor={blue!
50!
black}, urlcolor={blue!
80!
black}
}

\graphicspath{{figures/}}

\begin{document}

\maketitle

\begin{abstract}
Switchback experiments and other clustered randomized designs are widely used on online platforms, but the clustered, time-dependent nature of these designs can make standard variance reduction methods behave differently than in standard A/B tests.
We evaluate design-aware variance reduction methods for switchbacks---CUPED, CUPAC (ML-based covariate adjustment), and doubly robust (DR) estimators---relative to a baseline switchback analysis with cluster-robust standard errors.
Through a hierarchical simulation framework that varies key regime parameters---number of clusters, cluster-size imbalance, within-cluster autocorrelation, carryover, spillover, and predictive signal strength---we evaluate validity (false positive rate and confidence interval coverage) and efficiency (standard error reduction, power, and minimum detectable effect as a function of run length).
We also study how temporal carryover and spatial spillover affect bias and inference when SUTVA is violated.
The results summarize how the relative efficiency, finite-sample stability, and robustness of the four estimators shift across the design conditions we examine.
\end{abstract}

\section{Introduction}\label{sec:intro}

Switchback experiments randomize treatment assignments across time periods within geographic or operational clusters.
Unlike standard A/B tests, which randomize individual users, switchback designs are necessary when the intervention affects the entire marketplace within a cluster---for example, changes to pricing algorithms, dispatch logic, or matching algorithms in ride-hailing and other two-sided marketplace platforms \citep{chamandy2016experimentation, bajari2023experimental}.
The design has seen rapid adoption in the technology industry, where marketplace-level interventions cannot be cleanly randomized at the user level due to supply-side interactions \citep{johari2022experimental, bojinov2023design}.

A central challenge in switchback experiments is statistical power.
Because randomization occurs at the cluster--time-period level rather than the individual level, the number of independent units available for inference is far smaller than the number of observations.
The standard remedy is cluster-robust inference \citep{liang1986longitudinal, cameron2015practitioners}, which adjusts the standard errors to allow observations in the same cluster to move together in unrestricted ways.
However, when the number of clusters is modest, this correction is itself imprecise, because it relies on approximations that become accurate only as the number of clusters grows large.
The consequence is wide confidence intervals and, worse, intervals that contain the true effect less often than their stated 95\% confidence level promises.
This motivates variance reduction (VR) methods, which use covariates---variables that help predict the outcome but are themselves unaffected by the treatment---to subtract off predictable variation, narrowing confidence intervals without changing the quantity being estimated.

Three VR approaches dominate practice.
CUPED \citep{deng2013improving} adjusts outcomes using pre-experiment values of the outcome variable, typically a lagged average of the metric.\footnote{
    We adopt the following terminology throughout.
    \textit{Outcome} and \textit{metric} are used interchangeably for the dependent variable on which the treatment effect is estimated.
    A \textit{cluster} is a geographic cross-sectional unit (e.g., a city region), and a \textit{cell} is a cluster-by-time-period combination that serves as the unit of randomization.
    We describe clusters as geographic units throughout, as is conventional in marketplace switchbacks, but nothing in the analysis depends on this interpretation: the results apply to any partition of the population into cross-sectional units, for instance, groups of users, individual restaurants, or delivery vehicles.
}
CUPAC \citep{poyarkov2016boosted, li2024cupac} replaces the pre-period average with a real-time machine learning prediction, which frequently explains a larger share of the variation in the outcome and therefore achieves greater variance reduction.
Doubly robust estimation with double machine learning (DML DR) relies on two fitted components: an outcome model that predicts the metric from covariates, as in CUPAC, and a propensity model that predicts the probability that each unit is assigned to treatment.
The two are combined through an augmented inverse probability weighting (AIPW) correction \citep{robins1994estimation, chernozhukov2018double}, and each model is fitted on one portion of the data and applied to another so that errors made in fitting do not contaminate the treatment effect estimate.
The estimator recovers the true effect if either of the two models is correctly specified, which is the sense in which it is doubly robust, and it is most efficient when both are.

These methods have been primarily developed and validated in the standard A/B testing context with independent user-level randomization \citep{kohavi2020trustworthy, staponaite2025variance}.
Their comparative performance in a switchback setting, however, has not been examined systematically.
In switchbacks, severe cluster-size imbalance can drastically reduce statistical power by inflating the contribution of shocks that move an entire cluster or an entire time period at once \citep{pankratev2026powerful}.
Moreover, practitioners currently lack systematic guidance on how these estimators perform when a design is unfavorable---for example, when clusters are few or wildly unequal in size---and when the stable unit treatment value assumption (SUTVA) fails.
That assumption requires each unit's outcome to depend only on its own treatment assignment and not on the assignments given to other units.
Such violations are of particular concern in switchbacks, where randomization units are both larger and far scarcer, leaving fewer units over which contamination can average out.
To address this gap, we evaluate and compare these methods across several key dimensions that characterize real-world platform experiments: (i) experiment size (both in terms of the number of clusters and duration), (ii) cluster-size imbalance, (iii) covariate signal strength, which covers both the predictive accuracy of the machine learning model and the degree to which an outcome in one hour resembles the outcome in the hour before, and (iv) robustness to SUTVA violations arising from carryover over time and spillover across space.
Across these dimensions, we assess both the validity of the estimators, measured by bias and by how often the confidence interval contains the true effect, and their efficiency, measured by the reduction in the standard error and the resulting statistical power.\footnote{
    False positive rates are evaluated in the baseline regime by simulating a world in which the treatment has no effect at all and recording how often each method nonetheless reports a significant result.
    The minimum detectable effect, the smallest true effect an experiment could reliably detect, is reported for the baseline and experiment-duration analyses.
}

We develop a data-generating process (DGP) that models individual-level outcomes as a function of persistent differences between clusters, time-of-day effects, cluster-specific hourly fluctuations that persist across consecutive hours, treatment effects that vary across clusters, and optional contamination from carryover and spillover.
We evaluate four estimators---Raw (ordinary least squares, or OLS, with clustered standard errors), CUPED, CUPAC, and DML DR---across 26 distinct combinations of simulation settings, which we refer to as regimes, spanning seven design dimensions: the number of clusters, cluster-size imbalance, temporal autocorrelation, carryover intensity, spillover intensity, covariate signal strength, and experiment duration.
We run 500 simulations for the baseline parameter regimes and 200 simulations for all other regimes, producing 23{,}200 total estimator evaluations.

We interpret these results through the multi-level variance decomposition for switchback designs of \citet{pankratev2026powerful}, which we summarize in Section~\ref{sec:theory}.
The decomposition shows that the variance of the treatment effect estimator is governed by macro-level shocks---cluster, time, and cluster--time interaction effects---rather than by individual-level noise, because averaging over many individuals within a cell cancels out individual-level randomness, while unequal cluster sizes magnify the influence of the shocks that remain.
Under our baseline calibration, these macro shocks account for more than 99.5\% of the estimator variance even though they explain only about 28\% of the raw variation among individual observations.

Our simulation framework yields four main conclusions regarding the application of VR in switchback designs.
First, when the assumptions behind the adjustment hold and no interference is present, covariate adjustment yields substantial efficiency gains.
We summarize efficiency throughout by the standard error ratio, defined as the standard error of an adjusted method divided by the standard error of the unadjusted difference in means, so that smaller values indicate a more precise estimate.
CUPAC reduces this ratio to 0.50, equivalent to removing 75\% of the variance, and DML DR reduces it to 0.46, equivalent to removing 79\%, whereas CUPED achieves a considerably more modest ratio of 0.89, or a 20\% reduction.\footnote{
    In the absence of interference, all four methods produce unbiased treatment effect estimates, and their confidence intervals contain the true effect 95\% of the time, as intended.
    The differences between methods are purely in efficiency.
}

Second, power naturally increases with experiment size, but the estimators do not benefit equally.
Increasing the number of geographic units raises power for all estimators, but the two machine learning methods outperform the other two only once a sufficient number of clusters is available: with one thousand clusters, DML DR and CUPAC reach power of 0.96 and 0.91, respectively, compared with 0.49 for CUPED and 0.42 for Raw.
Extending the experiment along the time dimension has a comparable effect; under our baseline calibration, DML DR reaches 80\% power in approximately three days of experimentation, compared to two weeks for Raw.

With respect to covariate signal strength, the efficiency of DML DR and CUPAC is governed by the predictive power of the fitted real-time covariate, while CUPED instead draws on temporal persistence that makes lagged pre-period outcomes informative.\footnote{
    In our simulations the two adjustments are deliberately assigned distinct information sets: CUPAC, along with the outcome model underlying DML DR, exploits only real-time correlates of the outcome, whereas CUPED adjusts on pre-period outcome values.
    We separate the two information sets in order to isolate the distinct sources of predictive signal.
    In practice, a CUPAC model would typically be trained on both pre-period outcomes and real-time covariates, and would therefore also inherit the autocorrelation-driven gains that we attribute to CUPED here.
}
Raising the fraction of outcome variation explained by the CUPAC covariate from 15\% to 75\% lowers its standard error ratio from 0.85 to 0.26 and increases power from 0.17 to 0.88.
CUPED is insensitive to this dimension of signal, and responds instead to persistence over time: raising the correlation between consecutive hourly outcomes from zero to 0.9 improves its standard error ratio from 0.97 to 0.62 and more than doubles its power, from 0.12 to 0.31.
Even under such strong autocorrelation, however, CUPED remains substantially less efficient than both machine learning methods.

Third, design pathologies (severe cluster-size imbalance and small cluster counts) affect these estimators differentially.
Severe cluster-size imbalance (a coefficient of variation of 3.0 across cluster sizes) depresses power for every estimator---CUPAC falls from 0.67 to 0.26 and DML DR from 0.75 to 0.30 relative to a balanced design---exactly as the imbalance penalty derived in Section~\ref{sec:theory} predicts.
CUPAC's standard error ratio is essentially unaffected by imbalance, remaining near 0.50 throughout, whereas DML DR's rises modestly under severe imbalance, from 0.46 to 0.48, because the weights that correct for chance imbalance in assignment are themselves estimated imprecisely when clusters are small.

Small cluster counts present a distinct difficulty: with ten or fewer clusters, the confidence intervals of all four methods contain the true effect only about 82\% of the time rather than the intended 95\%, reflecting the well-documented tendency of cluster-robust standard errors to be too small when clusters are few rather than any property of the adjustments themselves.
CUPAC's standard error ratio remains near 0.50 even with as few as ten clusters, because it relies solely on a continuous covariate and has no propensity model to overfit.
DML DR, by contrast, is fragile in this regime: with few clusters, its propensity model overfits, raising its standard error ratio from 0.46 to 0.61.
The ordering between the two machine learning methods therefore reverses, since DML DR is more efficient than CUPAC when clusters are plentiful but less efficient than CUPAC when they are scarce.

Finally, we show that these efficiency gains become a liability when interference is severe.
We design our data-generating process so that carryover propagates across three lagged time periods, and we calibrate its intensity so that, at the upper end of the range we consider, the resulting bias can exceed the true treatment effect and reverse its sign.\footnote{
    The specific range of carryover intensities we simulate is chosen for illustrative purposes, so that the progression from mild attenuation to a full sign reversal is clearly visible in the results.
    The spillover mechanism follows an analogous logic and produces contamination of a comparable magnitude.
}
Such sign reversals are not merely a modeling device: \citet{xiong2023data} estimate 890 cumulative effect curves from real experiments on a ride-hailing platform and find that 68\% of them change sign as the treatment duration increases, indicating that non-monotonic, sign-reversing carryover is common in practice.
Under mild interference, all four methods suffer comparable bias, and the VR methods retain their power advantage, so they detect an effect that is biased but has the correct sign.
Under severe interference where bias exceeds the true treatment effect, the estimated effect therefore carries the opposite sign from the truth.
Greater precision is harmful in this situation, because a method with narrower confidence intervals is more likely to declare the incorrect sign statistically significant (a wrong-sign rejection).
Under severe carryover in our simulations, DML DR rejects in the wrong direction 38\% of the time, compared with 11\% for Raw, and spatial spillover produces the same pattern.
This is a specific instance of the classical efficiency--robustness tradeoff \citep{huber1964robust, hampel1986robust}: methods optimized for precision under one model become vulnerable when that model is misspecified.

Across the scenarios we study, CUPAC combines efficiency close to that of DML DR with substantially greater robustness.
It captures approximately 95\% of DML DR's efficiency gains under correct specification.
It matches or exceeds DML DR's stability under every design pathology we examine, because it relies on no propensity model that can overfit.
It also produces substantially fewer wrong-sign rejections than DML DR under severe interference.
Raw sits at the opposite extreme: it forgoes essentially all of the available efficiency gains, but for that reason it also produces the fewest wrong-sign rejections of any estimator when interference is severe.

This paper draws on and contributes to three literatures.
First, it contributes to the design and analysis of switchback experiments \citep{bojinov2019time, bojinov2023design, xiong2023data, hu2022switchback}.
That literature has concentrated on estimating treatment effects and on drawing inferences from the randomization itself when treatment carries over between periods, but it has not examined how variance reduction methods behave when the design is unfavorable.
Second, it extends the variance reduction literature for online experiments \citep{deng2013improving, li2024cupac, kohavi2020trustworthy} from the A/B testing setting to switchbacks, where clustering, temporal dependence, and interference create qualitatively new tradeoffs.
\citet{staponaite2025variance} conduct the comparison closest to ours, in standard A/B tests with independent user-level randomization.
They find that CUPAC performs best, narrowing confidence intervals by 38.5\%, whereas doubly robust estimation achieves a more modest reduction of approximately 21\%.
Doubly robust estimation performs considerably better in our switchback setting, reducing variance by 79\%, and the difference is attributable to the propensity component.
When only a few hundred units are randomized, the realized fraction assigned to treatment departs from the intended one-half by chance, and the propensity model corrects for the resulting imbalance.
When randomization instead occurs independently across a very large number of users, this chance imbalance is negligible and the propensity correction has little left to contribute.
Beyond this comparison, our switchback setting additionally introduces cluster structure and temporal dependence, and lets us evaluate all four estimators under SUTVA violations from carryover and spillover.
Third, it illustrates a concrete instance of the efficiency--robustness tradeoff from robust statistics \citep{huber1964robust, hampel1986robust}, connecting the abstract principle to a practical experimental design question faced by technology platforms.

The remainder of the paper is organized as follows.
Section~\ref{sec:setup} describes the simulation design, covering the data-generating process, the theoretical background on estimator variance and the imbalance penalty, and the four estimators we evaluate.
Section~\ref{sec:results} reports simulation results under the baseline regime, across design dimensions, and under carryover and spillover interference.
Section~\ref{sec:guidance} summarizes relative performance across experimental conditions.
Section~\ref{sec:conclusion} concludes.

\section{Simulation Design}\label{sec:setup}

We evaluate VR methods through Monte Carlo simulation of a switchback experiment.
We start by describing the data-generating process underlying the simulated outcomes (Section~\ref{sec:dgp}), then establish a theoretical result that determines which sources of variation a variance reduction method needs to capture in order to be effective (Section~\ref{sec:theory}), and finally introduce the four estimators we evaluate against that benchmark (Section~\ref{sec:estimators}).
The theoretical result plays a dual role in what follows: it guides how we calibrate the data-generating process, and it sets an a priori expectation for how much efficiency gain each variance reduction method's covariates should be able to deliver.

\subsection{Data-Generating Process}\label{sec:dgp}

The data-generating process (DGP) produces individual-level outcomes in a panel indexed by cluster $j = 1, \ldots, n_\text{cl}$, hour $h = 1, \ldots, n_t$, and individual observation $i = 1, \ldots, n_{jh}$.
We formally define the DGP using the potential outcomes framework \citep{rubin1980randomization}.
Let $Y_{ijh}(0)$ denote the potential outcome under control, defined as:
\begin{equation}\label{eq:y0}
    Y_{ijh}(0) = \mu + \alpha_j + \gamma_h + \delta_{jh} + \varepsilon_{ijh},
\end{equation}
where $\mu$ is the grand mean, $\alpha_j \sim \mathcal{N}(0, \sigma_\alpha)$ is a cluster random effect, $\gamma_h$ is a deterministic hour effect, $\delta_{jh}$ is a cluster--hour interaction following an AR(1) process with lag-1 autocorrelation $\rho$, and $\varepsilon_{ijh}$ is individual-level noise (the dominant variance component).
The potential outcome under treatment is $Y_{ijh}(1) = Y_{ijh}(0) + \tau_j$, where $\tau_j \sim \mathcal{N}(\tau, \tau_\text{sd}^2)$ is a heterogeneous treatment effect (HTE).
The observed outcome, allowing for interference via carryover ($C_{jh}$) and spillover ($S_{jh}$), is:
\begin{equation}\label{eq:dgp}
    Y_{ijh} = Y_{ijh}(T_{jh}) + C_{jh} + S_{jh} = Y_{ijh}(0) + \tau_j \cdot T_{jh} + C_{jh} + S_{jh},
\end{equation}
where $T_{jh} \sim \text{Bernoulli}(0.5)$ is the treatment indicator (constant within each cell).
When carryover and spillover are set to zero, Equation~\eqref{eq:dgp} satisfies SUTVA.
Cell sizes $n_{jh}$ are drawn from a log-normal--Poisson process with coefficient of variation CV, producing substantial cluster-size imbalance at the baseline $cv = 1.5$.
Specifically, we calibrate the baseline DGP using parameters representative of empirical data-generating processes commonly observed in practice.
The grand mean of the outcome variable $\mu$ is set to 2{,}000, with a total standard deviation $\sigma_\text{total}$ of 1{,}000.
The total variance is decomposed into shares of 5\% for the cluster random effects ($v_\alpha = 0.05$), 3\% for the deterministic hour effects ($v_\gamma = 0.03$), 20\% for the cluster--hour interaction effects ($v_\delta = 0.20$), and 72\% for the individual-level residual noise ($v_\varepsilon = 0.72$).
The mean cell size $\bar{n}$ is 180 observations per cluster-hour cell, and the treatment assignment probability is 0.5.
For the variance reduction methods, the CUPED covariate is calibrated to a pre-period target $R^2$ of 0.15, and the standard deviation of the heterogeneous treatment effects $\tau_\text{sd}$ is set to 10.
These baseline parameters, along with the ranges over which each parameter is varied in our design sensitivity analyses, are summarized in Table~\ref{tab:swept_ranges}, with a complete reference table provided in Appendix Table~\ref{tab:app_baseline_params}.

To model temporal carryover, we assume that when $\rho_C > 0$, treatment switches between consecutive periods produce contamination through a multi-lag mechanism:
\begin{equation}\label{eq:carryover}
    C_{jh} = \sum_{k=1}^{3} w_k \, \rho_C \, \tau_j \, (T_{j,h-k} - T_{jh}),
    \qquad w = (0.3,\, 0.2,\, 0.1).
\end{equation}
At $\rho_C \leq 1$, carryover attenuates the treatment effect toward zero; at $\rho_C > 1$, the cumulative bias exceeds the true effect and reverses the sign of the estimated average treatment effect (ATE).
Covariates are constructed from untreated potential outcomes and do not capture this contamination.

To model spatial spillover, we assume that when $\rho_S > 0$, treatment in nearby clusters leaks into control outcomes via the two nearest neighbors:\footnote{
    Specifically, we define proximity between clusters using their latent cluster random effects $\alpha_j$.
    For each cluster $j$, the first nearest neighbor $k_1$ is the cluster $k \neq j$ that minimizes $|\alpha_j - \alpha_k|$, and the second nearest neighbor $k_2$ is the cluster $k \neq j, k_1$ that minimizes $|\alpha_j - \alpha_k|$.
    The spillover intensity decays with distance in this latent space, with the nearest neighbor $k_1$ contributing full spillover and the second neighbor $k_2$ contributing half.
}
\begin{equation}\label{eq:spillover}
    S_{jh} = \rho_S \cdot (1 - T_{jh}) \cdot \bigl(\tau_{k_1} T_{k_1,h} + \tfrac{1}{2}\tau_{k_2} T_{k_2,h}\bigr).
\end{equation}
The $(1 - T_{jh})$ factor restricts spillover to control cells, creating attenuation bias.

Finally, we construct the covariates for our variance reduction methods as noisy proxies for the untreated potential outcome $Y^{(0)}$.
The CUPED covariate $x^\text{pre}$ has target $R^2 = 0.15$, constructed from the lagged interaction term $\delta_{j,h-1}$ and a partial sharing of the individual noise component.
The CUPAC covariate $x^\text{ml}$ has target $R^2 = 0.50$, constructed as $Y^{(0)}$ plus calibrated noise.

\begin{table}[t]
\centering
\caption{Simulation Regimes and Parameter Ranges}\label{tab:swept_ranges}
\renewcommand{\arraystretch}{1.25}
\resizebox{0.65\linewidth}{!}{%
\begin{tabular}{@{} l l l @{}}
    \toprule
    Dimension & Baseline Value & Values Examined \\
    \midrule
    Clusters $n_\text{cl}$              & 200   & $\{10,\, 50,\, 200,\, 500,\, 1{,}000\}$ \\
    Hours $n_t$                          & 24    & $\{12,\, 24,\, 48,\, 72,\, 168,\, 336\}$ \\
    Cluster-size CV                      & 1.5   & $\{0.5,\, 1.5,\, 3.0\}$ \\
    Autocorrelation $\rho$               & 0.3   & $\{0.0,\, 0.3,\, 0.6,\, 0.9\}$ \\
    CUPAC $R^2$                          & 0.50  & $\{0.15,\, 0.30,\, 0.50,\, 0.75\}$ \\
    Carryover $\rho_C$                   & 0.0   & $\{0.0,\, 0.5,\, 1,\, 2,\, 3\}$ \\
    Spillover $\rho_S$                   & 0.0   & $\{0.0,\, 0.1,\, 0.3,\, 0.5\}$ \\
    \bottomrule
\end{tabular}
}
\end{table}

We organize the simulations around seven design dimensions (Table~\ref{tab:swept_ranges}), each varied while holding others at baseline.
After deduplication, there are 26 unique parameter regimes.
The baseline regime is run under both the null ($\tau = 0$) and the alternative ($\tau = 20$) with 500 replications each; all other regimes use $\tau = 20$ with 200 replications, producing 23{,}200 total estimator evaluations.
Summary metrics recorded for each replication include: SE ratio ($\overline{\text{SE}}_\text{method} / \overline{\text{SE}}_\text{raw}$), variance reduction ($1 - \text{SE ratio}^2$), bias ($\overline{\hat{\tau}} - \tau$), coverage, power, and minimum detectable effect (MDE, calculated as $2.80 \times \overline{\text{SE}}$).

\subsection{Theoretical Background}\label{sec:theory}

Before examining the simulation results, we establish the theoretical limits of variance reduction in switchback designs.
The asymptotic variance of the unadjusted estimator under the DGP in \eqref{eq:y0} takes the following approximate form \citep{pankratev2026powerful}:
\begin{equation}\label{eq:variance}
    \text{Var}(\hat{\tau}) \approx \frac{4 \cdot \sigma_\text{total}^2}{n_\text{cl} \cdot n_t} \left[ S_\text{macro} \left( \frac{1}{\bar{n}} + 1 + cv^2 \right) + \frac{S_\text{res}}{\bar{n}} \right],
\end{equation}
where $\bar{n}$ is the mean cell size, $cv$ is the coefficient of variation of cluster sizes,\footnote{
    The coefficient of variation of cluster sizes is defined as $cv = \sigma_n / \bar{n}$, where $\sigma_n$ is the standard deviation of cluster sizes (measured as the number of observations per cluster) and $\bar{n}$ is the mean cluster size.
}
and $S_\text{macro}$ and $S_\text{res}$ represent the variance shares of the macro-level shocks ($\alpha_j$, $\gamma_h$, $\delta_{jh}$) and individual-level residual noise ($\varepsilon_{ijh}$), respectively.
Specifically, $S_\text{macro}$ is amplified by the imbalance penalty $(1 + cv^2)$, whereas $S_\text{res}$ is heavily attenuated by the cell size $\bar{n}$.
In dense markets where $\bar{n}$ is large, the individual-level residual noise term shrinks toward zero in the cluster-robust standard error, while the macro-level term does not vanish and is instead amplified by cluster-size imbalance.
As a result, how much of the macro-level shocks a covariate explains, rather than its overall predictive power over the raw outcome, is what determines its contribution to variance reduction.

To illustrate the magnitude of this effect, consider our baseline simulation parameters: mean cell size $\bar{n} = 180$ and cluster-size imbalance $cv = 1.5$.
Under the calibrated variance shares, the macro-level share (cluster, hour, and interaction effects combined) is $S_\text{macro} = 0.28$ and the individual residual share is $S_\text{res} = 0.72$.
Factoring out the common multiplier $4 \cdot \sigma_\text{total}^2 / (n_\text{cl} \cdot n_t)$, the variance contribution of individual-level residual noise is $S_\text{res} / \bar{n} = 0.72 / 180 = 0.004$.
In contrast, the variance contribution of the macro-level shocks is $S_\text{macro} (1/\bar{n} + 1 + cv^2) = 0.28 \times (1/180 + 1 + 2.25) \approx 0.912$.
Thus, despite macro-level shocks accounting for only 28\% of the raw individual-level noise, the cluster-size imbalance and the lack of complete averaging inflate their contribution so that they account for over 99.5\% of the total estimator variance ($0.912 / 0.916$).
This demonstrates that in dense, imbalanced markets, statistical power is almost entirely determined by macro-level shocks, making variance reduction targeting these shocks disproportionately valuable.

\subsection{Treatment Effect Estimators and Variance Reduction Methods}\label{sec:estimators}

All four estimators target the average treatment effect $\tau$ and produce a point estimate, a cluster-robust standard error \citep{liang1986longitudinal}, and a 95\% confidence interval; two-sided tests are conducted at $\alpha = 0.05$.

The first estimator we evaluate is the unadjusted difference-in-means (Raw), which is estimated via OLS of the outcome on the treatment indicator with cluster-robust standard errors.
This estimator uses no covariates and serves as our baseline benchmark.

The second estimator is CUPED \citep{deng2013improving}, which adjusts outcomes by projecting the target variable $Y$ onto a pre-experiment covariate $x^\text{pre}$ to remove predictable variance.
The adjustment acts as a control variate: $Y_{ijh}^\text{adj} = Y_{ijh} - \theta^\top(x_{ijh}^\text{pre} - \mathbb{E}[x^\text{pre}])$, where $\theta$ is the projection coefficient $\text{Var}(x^\text{pre})^{-1}\text{Cov}(x^\text{pre}, Y)$.
In finite samples, we estimate $\hat{\theta}$ via OLS and regress the empirical adjusted outcome $Y^\text{adj}$ on $T$ with cluster-robust standard errors.
The achievable variance reduction increases directly with the predictive power ($R^2$) of the covariate.

The third estimator is CUPAC \citep{poyarkov2016boosted, li2024cupac}, which is identical to CUPED but replaces the pre-period average $x^\text{pre}$ with an in-experiment machine learning prediction $x^\text{ml}$.
By projecting $Y$ onto a flexible ML prediction, CUPAC captures non-linear and high-dimensional relationships, yielding higher $R^2$.

The fourth estimator is DML DR \citep{chernozhukov2018double}, which is a doubly robust estimator based on the AIPW influence function \citep{robins1994estimation}.
Outcome and propensity nuisance models are estimated via two-fold cross-fitting at the cluster level; the ATE is then the mean of the AIPW influence function scores, with cluster-robust standard errors.\footnote{
    On each cross-fitting fold, outcome models $\hat{g}_t(x) = \text{OLS}(Y \sim x^\text{ml} \mid T = t)$ for $t \in \{0,1\}$ and a propensity model $\hat{e}(x) = \text{OLS}(T \sim x^\text{ml})$ are estimated.
    The AIPW influence function for observation $i$ is $\psi_i = \hat{g}_1(x_i) - \hat{g}_0(x_i) + T_i(Y_i - \hat{g}_1(x_i))/\hat{e}(x_i) - (1 - T_i)(Y_i - \hat{g}_0(x_i))/(1 - \hat{e}(x_i))$, and $\hat{\tau} = \bar{\psi}$.
}
The outcome model component is analogous to the CUPAC adjustment.
The propensity correction addresses a separate problem.
Although every cell is assigned to treatment with probability one-half, the fraction of cells actually assigned to treatment within any given cluster departs from one-half by chance, and clusters that happen to receive an unusually large or small share of treated cells distort a simple difference in means.
Weighting each observation by the inverse of its estimated assignment probability removes this chance imbalance.

\section{Simulation Results}\label{sec:results}

This section reports the results of the Monte Carlo simulations described in Section~\ref{sec:setup}.
We first establish a baseline benchmark under the default parameter regime with no interference (Section~\ref{sec:baseline}), then examine how relative efficiency and power change as experiment size, cluster-size imbalance, and covariate signal strength vary (Section~\ref{sec:design}), and finally study how temporal carryover and spatial spillover affect bias, standard errors, and inference (Section~\ref{sec:interference}).
Throughout, we compare the four estimators using standard error ratios, power, bias, and coverage.

\subsection{Baseline Efficiency}\label{sec:baseline}

\begin{figure}[t]
    \centering
    \includegraphics[width=\linewidth]{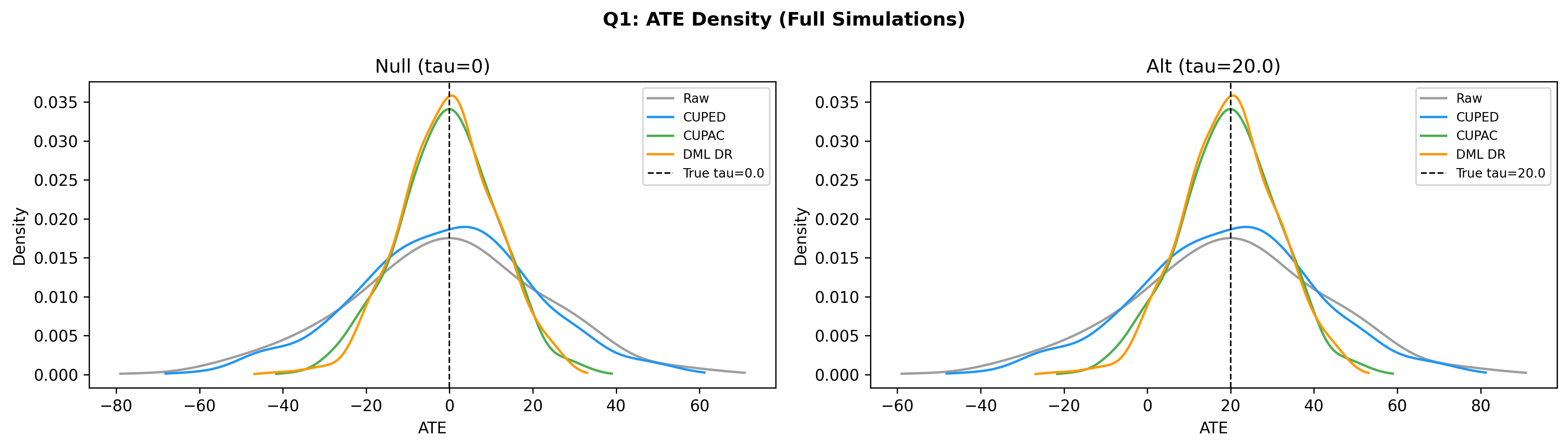}
    \caption{Distribution of ATE estimates across 500 replications under the alternative ($\tau = 20$).}\label{fig:q1_density}
\end{figure}

To evaluate the empirical performance of variance reduction methods in a switchback setting, we first establish a baseline benchmark under correct model specification and a standard, well-powered experimental design.
This baseline represents a typical experiment with $n_\text{cl} = 200$ clusters, a duration of $n_t = 24$ hours, baseline cluster-size imbalance ($cv = 1.5$), and moderate temporal autocorrelation ($\rho = 0.3$), with no carryover or spillover interference.
We run 500 independent Monte Carlo replications under both the null hypothesis ($\tau = 0$) to evaluate false positive rates and coverage, and the alternative hypothesis ($\tau = 20$) to evaluate power and MDE.
Table~\ref{tab:q1} reports the full suite of baseline performance metrics, and Figure~\ref{fig:q1_density} plots the empirical density of the treatment effect estimates across the 500 replications.

\begin{table}[t]
\centering
\caption{Baseline performance under the null and alternative hypotheses ($n_\text{cl} = 200$, $n_t = 24$, 500 replications).}\label{tab:q1}
\renewcommand{\arraystretch}{1.25}
\resizebox{\linewidth}{!}{%
\begin{tabular}{@{} l rrr rr c ccc @{}}
    \toprule
    Method & Bias & Mean SE & Emp.\ SE & SE Ratio & VR\% & Coverage & FPR & Power & MDE \\
    \midrule
    Raw   & $-0.85$ & 24.40 & 23.34 & 1.000 & 0.0\%  & 0.952 & 0.048 & 0.130 & 68.3 \\
    CUPED & $-0.87$ & 21.81 & 20.81 & 0.894 & 20.1\% & 0.958 & 0.042 & 0.136 & 61.1 \\
    CUPAC & $-0.44$ & 12.29 & 11.78 & 0.504 & 74.6\% & 0.956 & 0.044 & 0.372 & 34.4 \\
    DML DR   & $-0.26$ & 11.25 & 11.30 & 0.461 & 78.8\% & 0.956 & 0.044 & 0.450 & 31.5 \\
    \bottomrule
\end{tabular}}
\vspace{0.5ex}
{\small \textit{Note}: FPR = false positive rate ($\tau = 0$); Power = rejection rate ($\tau = 20$). MDE at 80\% power, two-sided $\alpha = 0.05$. SE Ratio relative to Raw.}
\end{table}

Moving from Raw to CUPAC reduces the standard error by roughly half (SE ratio 0.50, 75\% variance reduction), and DML DR reduces it further to an SE ratio of 0.46 (79\% variance reduction).
Given these reductions, power increases substantially: under our baseline calibration ($\tau = 20$), Raw achieves 13\% power while CUPAC reaches 37\% and DML DR reaches 45\%.
Under this calibration, DML DR's MDE of 31.5 is roughly half of Raw's 68.3.

All four methods achieve coverage close to the nominal 95\%.\footnote{
    This result depends on the use of cluster-robust standard errors.
    Because observations within a cluster share the cluster and cluster--hour effects, they are positively correlated, and standard errors computed as if observations were independent treat each observation as carrying more information than it does.
    Such standard errors understate the true sampling variability and produce coverage far below 95\%.
    Clustering the standard errors at the cluster level corrects this asymptotically, but confirming that the correction is adequate in finite samples---particularly for a semi-parametric estimator such as DML DR, which involves cross-fitting and an estimated propensity model---requires the direct verification we report here.
}
The close agreement between the mean estimated standard error (the average standard error computed within each simulation run) and the empirical standard error (the actual standard deviation of the point estimates across all 500 replications) indicates that the cluster-robust variance estimator is highly accurate in this baseline setting.
For DML DR, mean SE (11.25) and empirical SE (11.30) are in close agreement, consistent with well-behaved finite-sample performance at $n_\text{cl} = 200$.

The SE ratio, mean SE, empirical SE, and bias columns are identical across the $\tau = 0$ and $\tau = 20$ panels.
This is expected: the treatment coefficient absorbs the mean effect, while residual variance depends on the noise structure and HTE heterogeneity but not on $\tau$ itself.

All methods exhibit negligible bias (below 0.9 in absolute value, relative to standard errors of 11--24).
The efficiency gains are entirely attributable to the reduction in residual variance achieved by the covariate adjustment (except for DML DR, which additionally incorporates inverse probability weighting with an estimated propensity score).
Because the covariate is independent of the treatment assignment, the adjustment reduces the unexplained variance of the outcome without altering the expected value of the point estimator, thereby preserving its unbiasedness.

The efficiency gap between CUPED and CUPAC is driven by the predictive power of their respective covariates: CUPAC achieves the larger variance reduction in these simulations because it adjusts on a highly predictive machine learning model rather than on a simple pre-period lag.
CUPED's covariate $R^2$ of 0.15 and CUPAC's $R^2$ of 0.50 are calibrated to approximate typical industry settings, but actual predictive power will vary with the application and the ML model used.
We return to the relationship between covariate quality and efficiency in Section~\ref{sec:signal}.

\subsection{Design Sensitivity}\label{sec:design}

The baseline results establish that VR methods yield large efficiency gains under ideal conditions.
In practice, switchback experiments vary along several design dimensions that may interact with VR methods in non-obvious ways.
We organize these dimensions into three groups: experiment size (Section~\ref{sec:expsize}), cluster-size imbalance (Section~\ref{sec:q3}), and signal strength (Section~\ref{sec:signal}).
For each, we report SE ratios and power across the four methods, varying one dimension at a time while holding others at baseline.

\subsubsection{Experiment Size}\label{sec:expsize}

In this section, we analyze how the absolute and relative performance of variance reduction methods scales with the physical size of the experiment.
Experiment size in switchback designs has two distinct levers: the cross-sectional dimension (the number of clusters, $n_\text{cl}$) and the temporal dimension (the experiment duration, $n_t$).
While standard asymptotic theory suggests that standard errors for all estimators should scale proportionally to $1/\sqrt{n_\text{cl} \cdot n_t}$, this scaling may break down in finite samples for more complex estimators.
Specifically, we study how finite-sample constraints---such as a small number of clusters ($n_\text{cl} \leq 50$) or a short experiment duration (e.g., 12 to 24 hours)---interact with each estimator's performance.
We also evaluate how these two dimensions of experiment size determine the absolute time required to reach 80\% statistical power.

\begin{figure}[t]
    \centering
    \includegraphics[width=\linewidth]{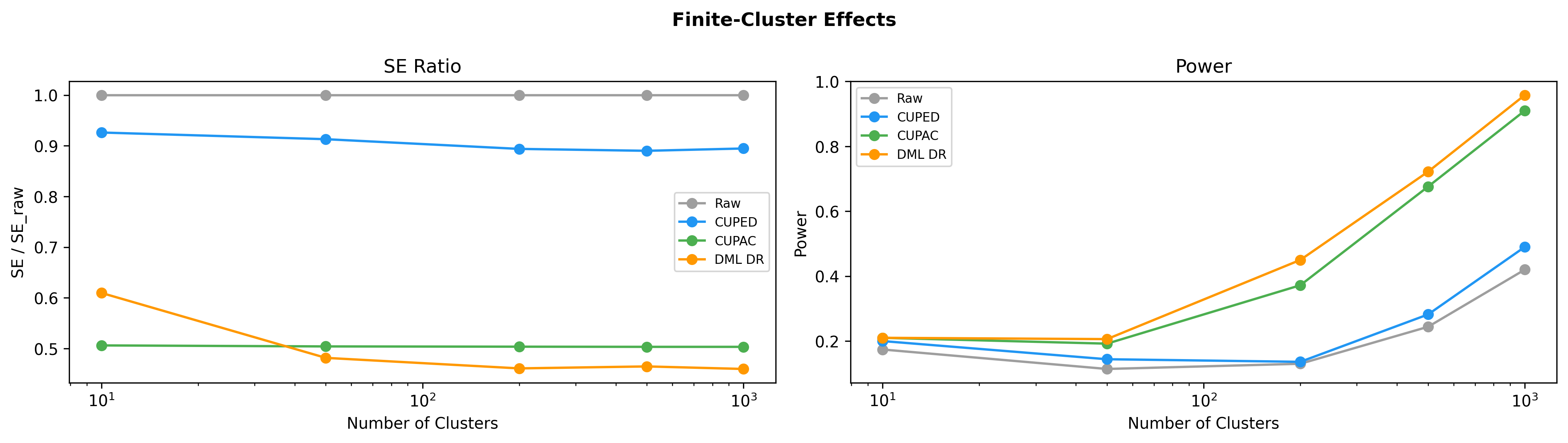}
    \caption{SE ratio and power as a function of the number of clusters ($n_\text{cl}$). 200 replications per regime, $\tau = 20$.}\label{fig:q2}
\end{figure}

First, we study how the number of experimental clusters ($n_\text{cl}$) affects the relative and absolute performance of each variance reduction method.
Figure~\ref{fig:q2} depicts the SE ratios and power across a range of 10 to 1{,}000 clusters (Appendix Table~\ref{tab:app_q2} presents the full numerical results).
SE ratios for Raw, CUPED, and CUPAC are roughly stable across $n_\text{cl}$: the covariate adjustment removes a fixed fraction of outcome variance regardless of the number of clusters.
DML DR, however, degrades at $n_\text{cl} = 10$, where its SE ratio rises to 0.61---exceeding CUPAC's 0.51 and substantially above its performance at larger cluster counts.
The mechanism is propensity model overfitting: with 10 clusters, the propensity model fits on the order of 33 parameters (cluster and hour dummies) from only 240 cell-level observations.
Cross-fitting with 5 clusters per fold produces noisy held-out propensity estimates, which inflate the inverse probability weighting (IPW) weights and increase the variance of the influence function $\psi$.
No such degradation occurs when the experiment is shortened along the time dimension instead, as we show below, because reducing the number of hours removes observations without adding parameters: the complexity of the propensity model is governed by the number of clusters, not by the number of time periods.
CUPAC avoids this problem because it uses only the continuous covariate and has no propensity model.

Power increases with $n_\text{cl}$ for all methods, as all standard errors scale proportionally to $1/\sqrt{n_\text{cl}}$.
A method attains high power once the true effect is roughly two standard errors or more in magnitude, so the methods with smaller standard errors---CUPAC and DML DR---reach that point at smaller cluster counts.
At $n_\text{cl} = 10$, coverage is below nominal for all methods (0.85--0.87), reflecting the well-known difficulty of cluster-robust inference with very few clusters \citep{cameron2015practitioners}.

\begin{figure}[t]
    \centering
    \includegraphics[width=\linewidth]{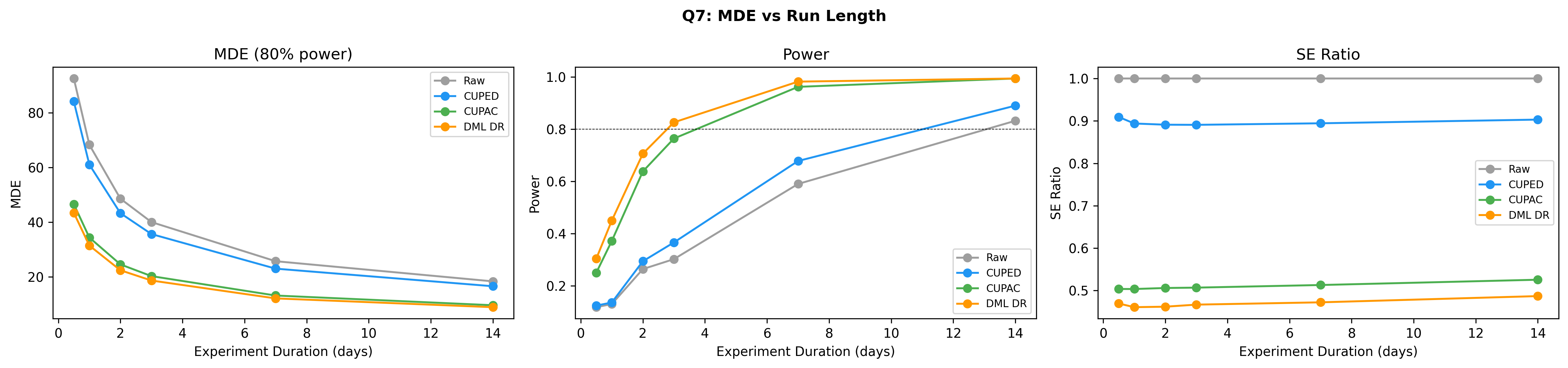}
    \caption{MDE, power, and SE ratio as a function of experiment duration. 200 replications per regime, $\tau = 20$.}\label{fig:q7}
\end{figure}

Second, we examine how the temporal dimension of the experiment---the overall duration ($n_t$)---interacts with the efficiency of each estimator.
Figure~\ref{fig:q7} displays the evolution of the minimum detectable effect, power, and SE ratio as the experiment duration increases from 12 hours (half a day) to 336 hours (two weeks; Appendix Table~\ref{tab:app_q7} presents the full numerical results).
VR methods provide the largest relative benefit at short durations.
Under our calibration, at 12 hours, DML DR's MDE is 43.4 compared to 92.5 for Raw---a factor of roughly 2.
DML DR reaches 80\% power in approximately three days, while Raw requires approximately two weeks to reach similar power levels.
CUPAC reaches 80\% power in approximately one week, compared to two weeks for Raw.

As duration increases, all methods converge toward full power and the relative advantage of VR narrows.
This is visible in the SE ratios: DML DR's SE ratio rises from 0.47 at 12 hours to 0.49 at 336 hours, and CUPAC's rises from 0.50 to 0.53.
The convergence occurs because additional hours add more data to all methods proportionally, while the covariate adjustment provides a fixed multiplicative reduction.
As the raw SE shrinks, the absolute gap between methods diminishes even though the proportional reduction is similar.

The practical implication is that VR methods matter most for short experiments---which is precisely when they are most needed, since short experiments are common for interventions that require rapid evaluation or carry high opportunity costs.

\subsubsection{Cluster-Size Imbalance}\label{sec:q3}

In this section, we examine how variance reduction methods interact with cluster-size imbalance---a pervasive pathology in online marketplaces where a few dominant, high-volume regions coexist with many smaller, low-volume ones.
In Section~\ref{sec:theory}, our theoretical framework showed that cluster-size imbalance inflates the variance contribution of macro-level shocks via the imbalance penalty $(1 + cv^2)$.
Here, we evaluate how this theoretical penalty translates into empirical power loss, and whether different VR methods can mitigate it.
Specifically, we study how increasing cluster-size imbalance affects the relative efficiency of outcome-adjusted versus propensity-adjusted estimators, and evaluate whether severe imbalance degrades the propensity score estimation in DML DR.
We also analyze the extent to which advanced variance reduction can salvage the statistical power lost to the imbalance penalty.

To understand the physical meaning of this imbalance penalty, it is instructive to examine how different values of the coefficient of variation (CV) translate into the actual distribution of cluster sizes.
In our simulations, the overall mean cell size $\bar{n}$ is held constant at 180 observations, but the latent cluster-level mean sizes are drawn from a log-normal distribution.
At a low imbalance level ($cv = 0.5$), the cluster sizes are relatively homogeneous: the median cluster size is approximately 160 observations, with the 10th and 90th percentiles spanning a narrow range of 100 to 280.
At our baseline imbalance level ($cv = 1.5$), which is calibrated to match typical online marketplace geographies, the distribution becomes highly right-skewed: the median cluster size drops to approximately 80 observations, while the 10th percentile is 25 and the 90th percentile is 420.
Under this baseline, the single largest cluster contains over 1{,}500 observations, meaning a single geographic region accounts for nearly 5\% of the entire experiment's volume.
At a severe imbalance level ($cv = 3.0$), the concentration becomes extreme: the median cluster size collapses to just 30 observations, while the single largest ``super-cluster'' can exceed 4{,}000 observations, dominating over 12\% of the entire experimental sample on its own.

\begin{figure}[t]
    \centering
    \includegraphics[width=\linewidth]{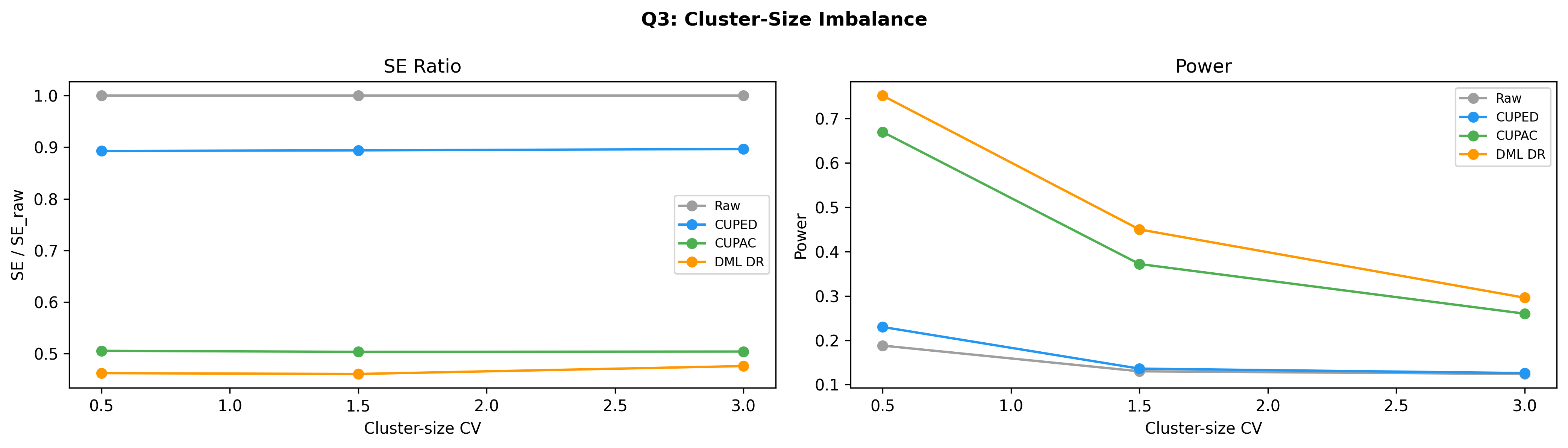}
    \caption{SE ratio and power as a function of cluster-size imbalance (CV). 200 replications per regime, $\tau = 20$.}\label{fig:q3}
\end{figure}

To study the empirical impact of cluster-size imbalance on the performance of each estimator, we run simulations across regimes where the coefficient of variation (CV) of cluster sizes ranges from 0.5 to 3.0.
Figure~\ref{fig:q3} displays the performance of each estimator across these different levels of imbalance, with the CV ranging from 0.5 (roughly equal clusters) to 3.0 (where a few large clusters dominate; Appendix Table~\ref{tab:app_q3} presents the full numerical results).
SE ratios are virtually constant across CV for Raw, CUPED, and CUPAC.
Imbalance inflates the standard error of the adjusted and the unadjusted estimator by the same proportional factor, because both are computed from the same observations and the same cluster weights, so the ratio between them is unaffected.
DML DR's SE ratio increases modestly at $cv = 3.0$ (from 0.46 to 0.48), because the propensity model's cluster dummies become poorly estimated for small clusters that contribute few observations.

The more notable pattern is in power.
Raw and CUPED experience a substantial decline in statistical power as imbalance increases (from 0.19 and 0.23 at $cv = 0.5$ to 0.12 and 0.13 at $cv = 3.0$), because they do not incorporate covariate adjustments to mitigate the increased estimator variance.
Consequently, the full proportional increase in the raw standard error driven by the imbalance penalty $(1 + cv^2)$ translates directly into power loss \citep{kish1965survey, eldridge2006sample, pankratev2026powerful}.
CUPAC and DML DR also decline substantially, from 0.67 to 0.26 and from 0.75 to 0.30, respectively.
Because their SE ratios are constant across CV, the absolute standard error grows with imbalance in the same proportion for every method, and the methods that begin from high power are the ones whose power has the furthest to fall.

\subsubsection{Signal Strength}\label{sec:signal}

Variance reduction methods rely fundamentally on the predictive signal of the covariates used for adjustment.
In a switchback setting, this predictive signal is determined by two distinct factors: the temporal structure of the underlying data-generating process (which dictates how much information historical lags carry about current outcomes) and the intrinsic predictive quality of the machine learning model (which determines the $R^2$ of the real-time covariate).
In this section, we systematically evaluate how these two dimensions of signal strength determine the efficiency of each method.
Specifically, we study how high temporal autocorrelation can allow a simple pre-period lag (CUPED) to capture temporal signals and approach the efficiency of a machine learning covariate (CUPAC).
We also analyze how the predictive power ($R^2$) of the outcome model determines the relative performance of CUPAC and DML DR, and examine whether there are diminishing returns to ML model optimization.

\begin{figure}[t]
    \centering
    \includegraphics[width=\linewidth]{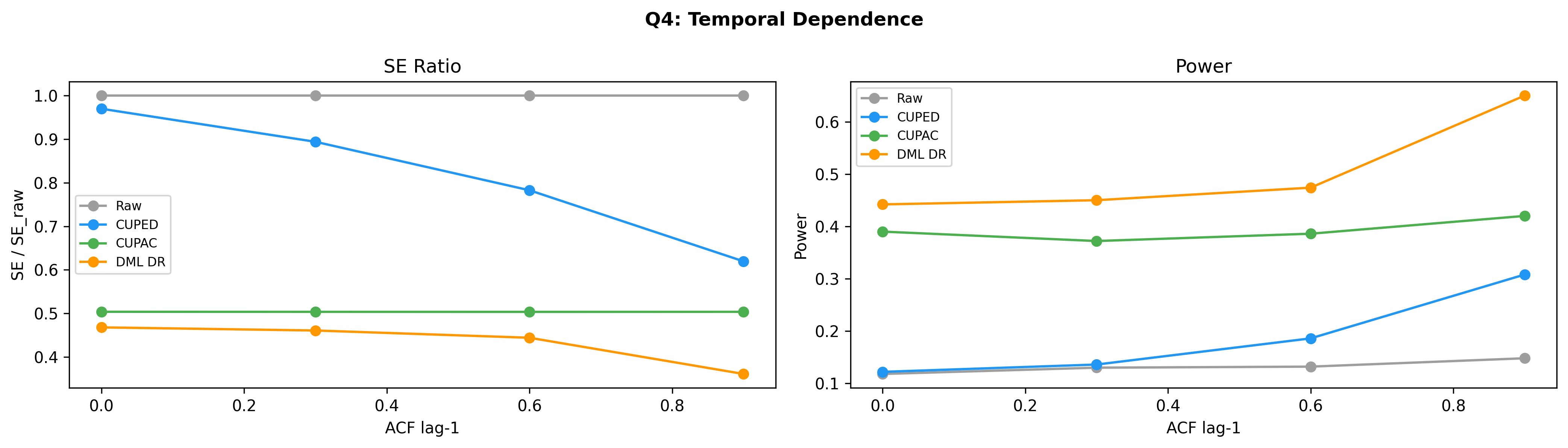}
    \caption{SE ratio and power as a function of lag-1 autocorrelation ($\rho$). 200 replications per regime, $\tau = 20$.}\label{fig:q4}
\end{figure}

First, we evaluate how the temporal autocorrelation ($\rho$) of the underlying outcome process determines the predictive value of lagged historical covariates.
Figure~\ref{fig:q4} presents the standard error ratios and power across different values of the lag-1 autocorrelation parameter $\rho$, ranging from 0.0 to 0.9 (Appendix Table~\ref{tab:app_q4} presents the full numerical results).
CUPED's efficiency improves substantially with higher autocorrelation.
Because the CUPED covariate includes a lagged interaction term $\delta_{j,h-1}$, its predictive power for the current outcome increases directly with $\rho$: at $\rho = 0.0$ the lagged term is uninformative and CUPED offers almost no improvement over Raw, while at $\rho = 0.9$ the lagged term is highly predictive and CUPED achieves SE reductions approaching those of CUPAC.

An important theoretical property of switchback designs is that temporal autocorrelation in the outcome process does not inflate the variance of the unadjusted estimator.
Because treatment assignments $T_{jh}$ are randomized independently across time periods, the covariance between outcomes in consecutive periods does not enter the variance of the difference-in-means estimator.
Consequently, the unadjusted Raw estimator's standard error and power remain virtually constant as autocorrelation increases.
However, autocorrelation directly determines the predictive efficiency of lagged covariates.
As autocorrelation increases, the lagged outcome becomes a much stronger predictor of the current outcome, allowing CUPED to capture this temporal signal and reduce the residual variance.

The SE ratios of CUPAC and DML DR are stable across values of $\rho$, because the CUPAC covariate is constructed with a fixed $R^2 = 0.50$ that does not depend on the temporal structure of the outcome.
DML DR shows a modest improvement at high $\rho$ (SE ratio from 0.46 at $\rho = 0.3$ to 0.36 at $\rho = 0.9$), suggesting that the more persistent cluster--hour interaction effects at high $\rho$ leave the propensity correction with more systematic cluster-level variation to remove.

In settings with strong temporal persistence---common in on-demand marketplaces and ride-hailing platforms, where demand conditions persist across consecutive hours---even a simple pre-period covariate can yield meaningful variance reduction.
The investment in a more elaborate ML covariate (CUPAC) is most valuable when temporal autocorrelation is low.

\begin{figure}[t]
    \centering
    \includegraphics[width=\linewidth]{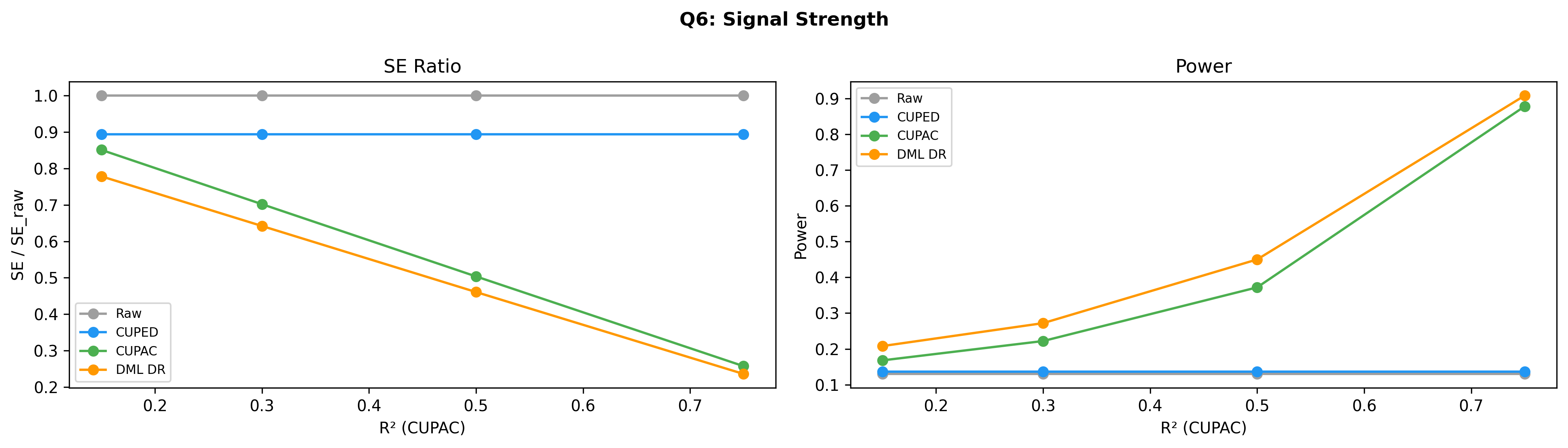}
    \caption{SE ratio and power as a function of CUPAC covariate $R^2$. CUPED $R^2$ is held fixed at 0.15. 200 replications per regime, $\tau = 20$.}\label{fig:q6}
\end{figure}

Second, we analyze how the intrinsic predictive quality of the machine learning model---measured by the covariate $R^2$---translates into empirical efficiency gains.
Figure~\ref{fig:q6} presents the SE ratios and power across different levels of the CUPAC covariate $R^2$, ranging from 0.15 (no better than CUPED) to 0.75 (an excellent ML model; Appendix Table~\ref{tab:app_q6} presents the full numerical results).
Raw and CUPED are unaffected, as expected---they do not use the ML covariate.

CUPAC's SE ratio decreases directly as its covariate $R^2$ increases: from 0.85 at $R^2 = 0.15$ (matching CUPED, as the covariates are equally predictive) to 0.26 at $R^2 = 0.75$.
DML DR also improves with $R^2$ but at a slower rate, moving from an SE ratio of 0.78 at $R^2 = 0.15$ to 0.24 at $R^2 = 0.75$.
The gap between CUPAC and DML DR narrows at higher $R^2$: at low $R^2$, the propensity correction in DML DR contributes a meaningful share of the total variance reduction, but at high $R^2$ the outcome model dominates and the propensity component becomes a smaller share of the total.
In other words, as the covariate improves, DML DR's additional complexity over CUPAC yields diminishing returns.

We note that our simulation isolates these estimators by design: the pre-experiment historical covariate used for CUPED is not included as a feature in the CUPAC model.
In practice, however, historical outcomes can and should be incorporated directly into the machine learning model's feature set.
In such cases, the machine learning model would natively capture the temporal autocorrelation signal, and the gap between CUPED and CUPAC observed under high autocorrelation would largely disappear.

\subsection{Randomization Interference}\label{sec:interference}

The preceding section examined how VR methods perform as design parameters vary, but under the maintained assumption that SUTVA holds.
We now relax this assumption and study the interaction between VR efficiency and two forms of interference: temporal carryover (Section~\ref{sec:q5}) and spatial spillover (Section~\ref{sec:q8}).
We evaluate the multi-lag carryover intensity across values of $\rho_C$ from 0 to 3, which translate directly to the relative magnitude of the resulting bias and determine whether the estimated treatment effect reverses sign.
Specifically, $\rho_C = 0.5$--$1$ in the DGP causes mild interference that attenuates the treatment effect toward zero, while $\rho_C = 2$--$3$ causes severe interference that produces a cumulative bias exceeding the true effect and flipping the sign of the estimated average treatment effect.
We demonstrate that while variance reduction methods significantly increase statistical power under correct specification, they amplify the risk of confidently rejecting in the wrong direction (Type~S error) when interference is severe enough to reverse the sign of the estimated treatment effect.

\subsubsection{Carryover}\label{sec:q5}

We begin our analysis of SUTVA violations by examining temporal carryover---a dynamic where the treatment assigned to a cluster in previous periods continues to affect outcomes in subsequent periods.
In marketplace settings, such carryover is common due to supply-side friction or persistent demand states.
Specifically, we study how the intensity of multi-lag carryover affects the bias and variance of both unadjusted and adjusted estimators.
We also evaluate whether using variance reduction covariates---which are constructed from untreated potential outcomes and do not account for active treatment contamination---exacerbates the risk of making a wrong-sign decision (Type~S error) when carryover is severe.
In general, carryover and spillover could also affect estimator variance by changing correlation structures or heteroskedasticity, but in our analysis this is a secondary concern: the DGP models both forms of contamination as additive mean shifts ($C_{jh}$ and $S_{jh}$ in Equations~\eqref{eq:carryover} and~\eqref{eq:spillover}) rather than as changes to the residual noise structure, so interference enters primarily through bias rather than through standard errors.

\begin{figure}[t]
    \centering
    \includegraphics[width=\linewidth]{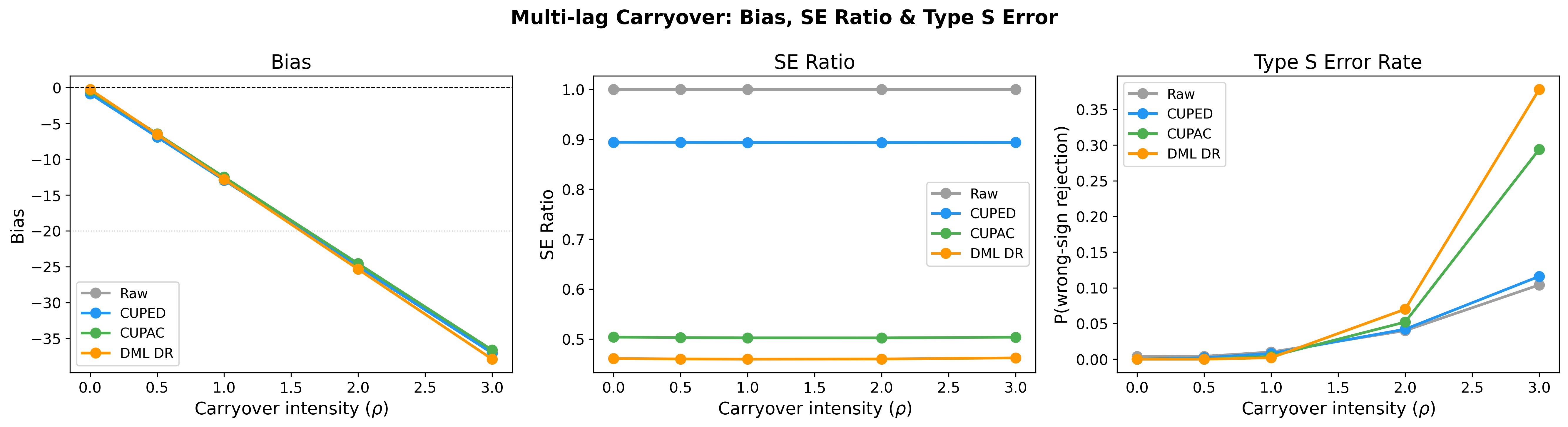}
    \caption{Bias, SE ratio, and wrong-sign rejection rate (Type~S error) as a function of multi-lag carryover intensity ($\rho_C$). 200 replications per regime, $\tau = 20$.}\label{fig:q5}
\end{figure}

Figure~\ref{fig:q5} shows the effect of increasing carryover intensity (which in our DGP spreads across multiple periods) from $\rho_C = 0$ (no carryover) to $\rho_C = 3$ (cumulative carryover nearly twice the treatment effect; Appendix Table~\ref{tab:app_q5} presents the full numerical results).
Two regimes emerge.

In the \textit{attenuating regime} ($\rho_C \leq 1$), carryover biases the estimated ATE toward zero but preserves its sign.
All methods suffer comparable bias---at $\rho_C = 1$, ranging from $-14.0$ (Raw) to $-14.8$ (DML DR)---because the contamination enters the DGP before any covariate adjustment takes place.
SE ratios are unaffected: carryover shifts outcomes systematically but does not alter the residual variance structure.
In this regime, VR methods retain their efficiency advantage, enabling detection of a biased but directionally correct effect.

In the \textit{sign-flip regime} ($\rho_C \geq 2$), the cumulative bias exceeds the true effect and the estimated ATE reverses sign.
At $\rho_C = 3$, the bias is approximately $-38$ to $-40$, so the estimated ATE is roughly $-18$ to $-20$---a strong negative signal when the true effect is $+20$.
VR methods, with their smaller standard errors, are now \textit{more likely} to detect this spurious negative effect as statistically significant.
In our simulation, DML DR achieves a wrong-sign rejection rate of 38\%, compared to 11\% for Raw.
CUPAC falls between them at 29\%.

This speaks directly to the classical efficiency--robustness tradeoff from robust statistics.
While optimal estimators maximize efficiency under correct specification, they often exhibit extreme vulnerability to small departures from model assumptions \citep{huber1964robust, hampel1986robust}.
In our setting, this tradeoff manifests as a vulnerability to interference: the same narrow confidence intervals that improve the detection of true treatment effects also substantially amplify the risk of confidently rejecting the null hypothesis in the wrong direction when SUTVA is violated.

\subsubsection{Spillover}\label{sec:q8}

Next, we turn to spatial spillover---a SUTVA violation where the treatment assigned to one cluster leaks into neighboring clusters.
In physical marketplaces, spatial spillovers may occur due to shared resources (such as drivers or couriers moving across cluster boundaries) or customer demand substitution.
We study whether spatial spillover produces different bias and variance trajectories compared to temporal carryover, and evaluate how increasing spillover intensity ($\rho_S$) affects bias, standard error ratios, and coverage across the four estimators.

\begin{figure}[t]
    \centering
    \includegraphics[width=\linewidth]{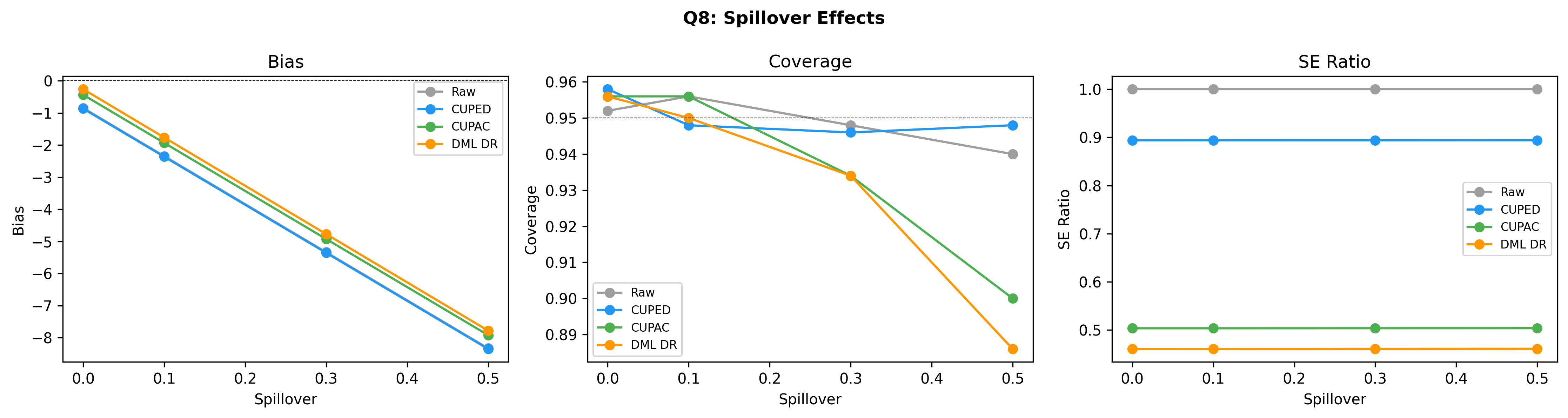}
    \caption{Bias, SE ratio, and wrong-sign rejection rate as a function of spillover intensity ($\rho_S$). 200 replications per regime, $\tau = 20$.}\label{fig:q8}
\end{figure}

Figure~\ref{fig:q8} shows the effect of increasing spillover intensity from $\rho_S = 0$ to $\rho_S = 0.5$ (Appendix Table~\ref{tab:app_q8} presents the full numerical results).
Within this range, spillover reproduces the qualitative pattern seen in the attenuating carryover regime ($\rho_C \leq 1$): all methods incur comparable negative bias that grows with $\rho_S$, because the contamination enters the DGP before any covariate adjustment takes place.
At $\rho_S = 0.5$, bias ranges from $-8.3$ (Raw) to $-7.8$ (DML DR), attenuating the estimated average treatment effect toward zero while preserving its sign.
SE ratios are essentially unchanged across spillover levels---CUPAC remains near 0.50 and DML DR near 0.46---because spillover shifts outcomes systematically without altering the residual variance structure that the covariate adjustment targets.

Unlike carryover, we do not observe a sign-flip regime within the spillover intensities examined.
Even at $\rho_S = 0.5$, the estimated ATE remains positive (roughly $+12$ when the true effect is $+20$), and wrong-sign rejection rates in Figure~\ref{fig:q8} remain near zero.
The efficiency--robustness tension nonetheless appears in coverage: at $\rho_S = 0.5$, coverage falls to 0.90 for CUPAC and 0.89 for DML DR, compared with 0.94 for Raw, because the same attenuating bias represents a larger share of a narrower confidence interval.

Due to how the DGP is constructed, spillover affects only control cells with treated neighbors, whereas carryover affects every cell that experiences a treatment flip (roughly half, under i.i.d. assignment), so bias accumulates more slowly as the spillover intensity parameter increases.
Within the spillover intensities we examine ($\rho_S \leq 0.5$), bias reaches approximately $-8$, whereas carryover at $\rho_C = 3$ produces bias of roughly $-38$ and enters the sign-flip regime.


\section{Relative Performance Summary}\label{sec:guidance}

This section synthesizes the empirical findings established across the various design dimensions, comparing how the four estimators' relative efficiency, finite-sample stability, and robustness to SUTVA violations shift across experimental conditions.
Table~\ref{tab:summary} summarizes which method performs best and worst along each dimension, and the discussion that follows describes how this comparison shifts with cluster count, temporal autocorrelation, interference severity, and implementation constraints.

\begin{table}[t]
\centering
\caption{Summary of method performance across design dimensions.}\label{tab:summary}
\renewcommand{\arraystretch}{1.30}
\resizebox{\linewidth}{!}{%
\begin{tabular}{@{} l l l @{}}
    \toprule
    Dimension & Best performer & Weakest performer \\
    \midrule
    Baseline efficiency           & DML DR  & Raw \\
    Small $n_\text{cl}$ ($\leq 50$) & CUPAC & DML DR \\
    High cluster imbalance        & CUPAC (stable SE ratio), DML DR (retains power) & Raw (steepest power loss) \\
    High autocorrelation          & CUPED, DML DR & Raw \\
    Mild interference (attenuating) & DML DR / CUPAC & --- \\
    Severe interference (sign-flip risk) & Raw & DML DR \\
    Short experiments             & DML DR  & Raw \\
    Implementation simplicity     & Raw / CUPED & DML DR \\
    \bottomrule
\end{tabular}}
\end{table}

No single method dominates across all dimensions: which method compares favorably depends on the experimental setting.
We describe how this comparison shifts along four dimensions: cluster count together with SUTVA plausibility, temporal autocorrelation, the severity of interference, and implementation constraints.

First, when SUTVA is plausible and the experiment involves at least 50 clusters, DML DR delivers the largest efficiency gains among the four estimators, provided randomization periods are long enough for transient effects to dissipate or clusters are geographically separated.
The cluster count requirement reflects the propensity model's need for sufficient data to avoid overfitting.
This efficiency advantage comes with substantially higher engineering complexity, however, since DML DR requires multi-fold cross-fitting and the estimation of both propensity and outcome models, which can be difficult to implement and maintain in production pipelines.

Furthermore, the temporal autocorrelation of the outcome process plays a critical role in determining the efficiency of both CUPED and DML DR.
In environments characterized by high temporal persistence, the historical pre-period covariate becomes highly predictive of current outcomes, allowing the simpler CUPED estimator to achieve substantial variance reduction.
Simultaneously, high autocorrelation increases the variance contribution of the cluster--hour interaction effects, making the cluster-level structure more exploitable by DML DR's propensity correction and further improving its relative efficiency.

Second, under mild interference (the attenuating regime, $\rho_C \leq 1$), both CUPAC and DML DR retain most of their efficiency advantage and enable detection of a biased but directionally correct effect.
CUPAC captures roughly 95\% of DML DR's variance reduction at baseline while avoiding propensity overfitting and remaining stable across cluster counts and imbalance levels, provided the covariate maintains predictive power ($R^2 \gtrsim 0.3$); DML DR's additional gain over CUPAC in this regime comes with the added complexity and small-sample sensitivity described above.

Third, when severe interference is possible (the sign-flip regime), unadjusted Raw estimation provides the lowest wrong-sign rejection rate.
In our simulation at $\rho_C = 3$, Raw rejected in the wrong direction only 11\% of the time, compared to 38\% for DML DR.
The unadjusted estimator may also serve as a diagnostic benchmark: substantial disagreement between VR-adjusted and unadjusted estimates can serve as an early warning of interference.

Finally, implementation constraints also shift which method compares favorably.
When experiments are small ($n_\text{cl} < 50$), CUPAC outperforms DML DR, which degrades at small cluster counts.
Without ML infrastructure, CUPED offers modest but reliable variance reduction with minimal implementation effort, and in settings with high temporal autocorrelation, CUPED's efficiency approaches that of CUPAC, narrowing the gap that would otherwise favor an ML covariate.

\section{Conclusion}\label{sec:conclusion}

This paper studies variance reduction methods for switchback experiments through Monte Carlo simulation across parameter regimes calibrated to real-world operational settings.
Our results confirm that when pre-experimental or real-time covariates have high predictive power, variance reduction substantially increases statistical power and can reduce the required length and size of experiments.
These efficiency gains have direct practical value: reaching adequate power in days rather than weeks allows a platform to evaluate and deploy a larger number of candidate interventions within a fixed period of time.

These gains are not uniform across designs.
Cluster-size imbalance depresses power for every method in the proportion that the imbalance penalty predicts, and very small cluster counts undermine cluster-robust inference for all four estimators alike.
Small cluster counts additionally penalize DML DR specifically, because its propensity model overfits when few clusters are available.

We also demonstrate that these same efficiency gains create a vulnerability under interference.
When interference is mild, variance reduction methods retain their power advantage and enable the detection of a biased but directionally correct treatment effect.
When interference is severe enough to reverse the sign of the estimated treatment effect, however, the narrower confidence intervals of these methods substantially increase the risk of confidently rejecting the null hypothesis in the wrong direction.
Taken together, these results show that CUPAC compares favorably across most of the conditions we study, since it obtains nearly all of the efficiency available to DML DR while remaining stable across design dimensions and less prone to wrong-sign errors, and they indicate that the unadjusted estimator retains value as a diagnostic benchmark whenever interference cannot be ruled out.

\section*{Acknowledgments}

We are grateful to the participants of the American Causal Inference Conference (ACIC) 2026 and the Symposium on Data Science and Statistics (SDSS) 2026 for their valuable comments and suggestions, which substantially improved this work.

\bibliographystyle{apalike}
\bibliography{references}

\clearpage
\appendix
\section{Supplementary Tables}\label{app:tables}

This appendix provides the full numerical results underlying the figures in Section~\ref{sec:results}.
All regimes use $\tau = 20$ with 200 replications except where noted.

\begin{table}[H]
\centering
\caption{Baseline Data-Generating Process (DGP) Parameters}\label{tab:app_baseline_params}
\renewcommand{\arraystretch}{1.25}
\begin{tabular}{@{} l l @{}}
    \toprule
    Parameter & Baseline Value \\
    \midrule
    Grand mean $\mu$                     & 2{,}000 \\
    Total std $\sigma_\text{total}$      & 1{,}000 \\
    Variance shares $(v_\alpha, v_\gamma, v_\delta, v_\varepsilon)$ & $(0.05,\, 0.03,\, 0.20,\, 0.72)$ \\
    Mean cell size $\bar{n}$             & 180 \\
    Treatment probability                & 0.5 \\
    CUPED $R^2$                          & 0.15 \\
    HTE std $\tau_\text{sd}$             & 10 \\
    \bottomrule
\end{tabular}
\end{table}

\subsection{Experiment Size}\label{app:expsize}

Table~\ref{tab:app_q2} and Table~\ref{tab:app_q7} present the detailed simulation results across different cluster counts and experiment durations, respectively.

\begin{table}[H]
\centering
\caption{Performance by number of clusters ($n_\text{cl}$), $\tau = 20$.}\label{tab:app_q2}
\renewcommand{\arraystretch}{1.25}
\footnotesize
\begin{tabular}{@{} r rrr rrr rrr rrr @{}}
    \toprule
    $n_\text{cl}$ & \multicolumn{3}{c}{Raw} & \multicolumn{3}{c}{CUPED} & \multicolumn{3}{c}{CUPAC} & \multicolumn{3}{c}{DML DR} \\
    \cmidrule(lr){2-4} \cmidrule(lr){5-7} \cmidrule(lr){8-10} \cmidrule(lr){11-13}
     & SE r. & Pwr & Cov. & SE r. & Pwr & Cov. & SE r. & Pwr & Cov. & SE r. & Pwr & Cov. \\
    \midrule
    10   & 1.000 & 0.17 & 0.82 & 0.926 & 0.20 & 0.81 & 0.506 & 0.21 & 0.82 & 0.610 & 0.21 & 0.83 \\
    50   & 1.000 & 0.11 & 0.92 & 0.913 & 0.14 & 0.91 & 0.504 & 0.19 & 0.93 & 0.482 & 0.21 & 0.91 \\
    200  & 1.000 & 0.13 & 0.95 & 0.894 & 0.14 & 0.96 & 0.504 & 0.37 & 0.96 & 0.461 & 0.45 & 0.96 \\
    500  & 1.000 & 0.24 & 0.94 & 0.890 & 0.28 & 0.94 & 0.503 & 0.68 & 0.94 & 0.465 & 0.72 & 0.93 \\
    1000 & 1.000 & 0.42 & 0.94 & 0.895 & 0.49 & 0.93 & 0.503 & 0.91 & 0.95 & 0.460 & 0.96 & 0.95 \\
    \bottomrule
\end{tabular}

\vspace{0.5ex}
{\small \textit{Note}: SE r.\ = SE ratio relative to Raw; Pwr = power; Cov.\ = 95\% CI coverage. 500 replications at $n_\text{cl} = 200$; 200 at others.}
\end{table}

\begin{table}[H]
\centering
\caption{MDE, power, and SE ratio by experiment duration, $\tau = 20$.}\label{tab:app_q7}
\renewcommand{\arraystretch}{1.25}
\resizebox{\linewidth}{!}{%
\begin{tabular}{@{} rr rrr rrr rrr rrr @{}}
    \toprule
    Hours & Days & \multicolumn{3}{c}{Raw} & \multicolumn{3}{c}{CUPED} & \multicolumn{3}{c}{CUPAC} & \multicolumn{3}{c}{DML DR} \\
    \cmidrule(lr){3-5} \cmidrule(lr){6-8} \cmidrule(lr){9-11} \cmidrule(lr){12-14}
     & & MDE & Pwr & SE r. & MDE & Pwr & SE r. & MDE & Pwr & SE r. & MDE & Pwr & SE r. \\
    \midrule
    12  & 0.5  & 92.5 & 0.12 & 1.000 & 84.1 & 0.12 & 0.909 & 46.6 & 0.25 & 0.504 & 43.4 & 0.30 & 0.469 \\
    24  & 1.0  & 68.3 & 0.13 & 1.000 & 61.1 & 0.14 & 0.894 & 34.4 & 0.37 & 0.504 & 31.5 & 0.45 & 0.461 \\
    48  & 2.0  & 48.6 & 0.26 & 1.000 & 43.3 & 0.29 & 0.891 & 24.6 & 0.64 & 0.506 & 22.5 & 0.71 & 0.462 \\
    72  & 3.0  & 40.1 & 0.30 & 1.000 & 35.7 & 0.37 & 0.891 & 20.3 & 0.76 & 0.507 & 18.7 & 0.83 & 0.467 \\
    168 & 7.0  & 25.8 & 0.59 & 1.000 & 23.0 & 0.68 & 0.894 & 13.2 & 0.96 & 0.513 & 12.2 & 0.98 & 0.472 \\
    336 & 14.0 & 18.4 & 0.83 & 1.000 & 16.6 & 0.89 & 0.903 &  9.7 & 0.99 & 0.526 &  9.0 & 0.99 & 0.487 \\
    \bottomrule
\end{tabular}}

\vspace{0.5ex}
{\small \textit{Note}: MDE $= (z_{0.025} + z_{0.20}) \times \text{SE}$; Pwr = power; SE r.\ = SE ratio relative to Raw at same duration. 500 replications at 24h; 200 at others.}
\end{table}

\subsection{Cluster-Size Imbalance}\label{app:q3}

Table~\ref{tab:app_q3} presents the detailed simulation results across different levels of cluster-size imbalance.

\begin{table}[H]
\centering
\caption{Performance by cluster-size coefficient of variation ($cv$), $\tau = 20$.}\label{tab:app_q3}
\renewcommand{\arraystretch}{1.25}
\footnotesize
\begin{tabular}{@{} r rrr rrr rrr rrr @{}}
    \toprule
    $cv$ & \multicolumn{3}{c}{Raw} & \multicolumn{3}{c}{CUPED} & \multicolumn{3}{c}{CUPAC} & \multicolumn{3}{c}{DML DR} \\
    \cmidrule(lr){2-4} \cmidrule(lr){5-7} \cmidrule(lr){8-10} \cmidrule(lr){11-13}
     & SE r. & Pwr & Cov. & SE r. & Pwr & Cov. & SE r. & Pwr & Cov. & SE r. & Pwr & Cov. \\
    \midrule
    0.5 & 1.000 & 0.19 & 0.95 & 0.893 & 0.23 & 0.95 & 0.506 & 0.67 & 0.96 & 0.462 & 0.75 & 0.95 \\
    1.5 & 1.000 & 0.13 & 0.95 & 0.894 & 0.14 & 0.96 & 0.504 & 0.37 & 0.96 & 0.461 & 0.45 & 0.96 \\
    3.0 & 1.000 & 0.12 & 0.94 & 0.896 & 0.13 & 0.92 & 0.504 & 0.26 & 0.94 & 0.476 & 0.30 & 0.93 \\
    \bottomrule
\end{tabular}

\vspace{0.5ex}
{\small \textit{Note}: SE r.\ = SE ratio relative to Raw; Pwr = power; Cov.\ = 95\% CI coverage. 500 replications at $cv = 1.5$; 200 at others.}
\end{table}

\subsection{Signal Strength}\label{app:signal}

Table~\ref{tab:app_q4} and Table~\ref{tab:app_q6} present the detailed simulation results across different levels of temporal autocorrelation and CUPAC covariate $R^2$, respectively.

\begin{table}[H]
\centering
\caption{Performance by lag-1 autocorrelation ($\rho$), $\tau = 20$.}\label{tab:app_q4}
\renewcommand{\arraystretch}{1.25}
\footnotesize
\begin{tabular}{@{} r rrr rrr rrr rrr @{}}
    \toprule
    $\rho$ & \multicolumn{3}{c}{Raw} & \multicolumn{3}{c}{CUPED} & \multicolumn{3}{c}{CUPAC} & \multicolumn{3}{c}{DML DR} \\
    \cmidrule(lr){2-4} \cmidrule(lr){5-7} \cmidrule(lr){8-10} \cmidrule(lr){11-13}
     & SE r. & Pwr & Cov. & SE r. & Pwr & Cov. & SE r. & Pwr & Cov. & SE r. & Pwr & Cov. \\
    \midrule
    0.0 & 1.000 & 0.12 & 0.96 & 0.970 & 0.12 & 0.95 & 0.504 & 0.39 & 0.95 & 0.468 & 0.44 & 0.96 \\
    0.3 & 1.000 & 0.13 & 0.95 & 0.894 & 0.14 & 0.96 & 0.504 & 0.37 & 0.96 & 0.461 & 0.45 & 0.96 \\
    0.6 & 1.000 & 0.13 & 0.95 & 0.783 & 0.19 & 0.96 & 0.504 & 0.39 & 0.94 & 0.444 & 0.47 & 0.95 \\
    0.9 & 1.000 & 0.15 & 0.94 & 0.620 & 0.31 & 0.96 & 0.504 & 0.42 & 0.94 & 0.361 & 0.65 & 0.95 \\
    \bottomrule
\end{tabular}

\vspace{0.5ex}
{\small \textit{Note}: SE r.\ = SE ratio relative to Raw; Pwr = power; Cov.\ = 95\% CI coverage. 500 replications at $\rho = 0.3$; 200 at others.}
\end{table}

\begin{table}[H]
\centering
\caption{Performance by CUPAC covariate $R^2$, $\tau = 20$. CUPED $R^2$ is held fixed at 0.15.}\label{tab:app_q6}
\renewcommand{\arraystretch}{1.25}
\footnotesize
\begin{tabular}{@{} r rrr rrr rrr rrr @{}}
    \toprule
    $R^2$ & \multicolumn{3}{c}{Raw} & \multicolumn{3}{c}{CUPED} & \multicolumn{3}{c}{CUPAC} & \multicolumn{3}{c}{DML DR} \\
    \cmidrule(lr){2-4} \cmidrule(lr){5-7} \cmidrule(lr){8-10} \cmidrule(lr){11-13}
     & SE r. & Pwr & Cov. & SE r. & Pwr & Cov. & SE r. & Pwr & Cov. & SE r. & Pwr & Cov. \\
    \midrule
    0.15 & 1.000 & 0.13 & 0.95 & 0.894 & 0.14 & 0.96 & 0.851 & 0.17 & 0.95 & 0.778 & 0.21 & 0.96 \\
    0.30 & 1.000 & 0.13 & 0.95 & 0.894 & 0.14 & 0.96 & 0.702 & 0.22 & 0.96 & 0.642 & 0.27 & 0.96 \\
    0.50 & 1.000 & 0.13 & 0.95 & 0.894 & 0.14 & 0.96 & 0.504 & 0.37 & 0.96 & 0.461 & 0.45 & 0.96 \\
    0.75 & 1.000 & 0.13 & 0.95 & 0.894 & 0.14 & 0.96 & 0.257 & 0.88 & 0.95 & 0.236 & 0.91 & 0.96 \\
    \bottomrule
\end{tabular}

\vspace{0.5ex}
{\small \textit{Note}: SE r.\ = SE ratio relative to Raw; Pwr = power; Cov.\ = 95\% CI coverage. 200 replications per regime.}
\end{table}

\subsection{Interference}\label{app:interference}

Table~\ref{tab:app_q5} and Table~\ref{tab:app_q8} present the detailed simulation results for temporal carryover and spatial spillover, respectively.

\begin{table}[H]
\centering
\caption{Multi-lag carryover effects ($\tau = 20$).}\label{tab:app_q5}
\renewcommand{\arraystretch}{1.25}
\resizebox{\linewidth}{!}{%
\begin{tabular}{@{} r rrr rrr rrr rrr @{}}
    \toprule
    $\rho_C$ & \multicolumn{3}{c}{Raw} & \multicolumn{3}{c}{CUPED} & \multicolumn{3}{c}{CUPAC} & \multicolumn{3}{c}{DML DR} \\
    \cmidrule(lr){2-4} \cmidrule(lr){5-7} \cmidrule(lr){8-10} \cmidrule(lr){11-13}
     & Bias & SE r. & Pwr & Bias & SE r. & Pwr & Bias & SE r. & Pwr & Bias & SE r. & Pwr \\
    \midrule
    0   & $-0.2$  & 1.00 & 0.13 & $-0.2$  & 0.89 & 0.14 & $-0.1$  & 0.50 & 0.37 & $+0.0$  & 0.46 & 0.45 \\
    0.5 & $-7.9$  & 1.00 & 0.08 & $-8.3$  & 0.89 & 0.08 & $-6.9$  & 0.50 & 0.21 & $-8.4$  & 0.46 & 0.25 \\
    1   & $-14.0$ & 1.00 & 0.05 & $-14.4$ & 0.89 & 0.05 & $-13.0$ & 0.50 & 0.09 & $-14.8$ & 0.46 & 0.11 \\
    2   & $-26.2$ & 1.00 & 0.05 & $-26.6$ & 0.89 & 0.05 & $-25.2$ & 0.50 & 0.06 & $-27.5$ & 0.46 & 0.07 \\
    3   & $-38.4$ & 1.00 & 0.11 & $-38.8$ & 0.89 & 0.12 & $-37.4$ & 0.50 & 0.29 & $-40.2$ & 0.46 & 0.38 \\
    \bottomrule
\end{tabular}}

\vspace{0.5ex}
{\small \textit{Note}: SE r.\ = SE ratio relative to Raw; Pwr = two-sided rejection rate. At $\rho_C \geq 2$, Pwr denotes wrong-sign rejections (Type~S error). 500 replications at $\rho_C = 0$; 200 at others.}
\end{table}

\begin{table}[H]
\centering
\caption{Spatial spillover effects ($\tau = 20$).}\label{tab:app_q8}
\renewcommand{\arraystretch}{1.25}
\resizebox{\linewidth}{!}{%
\begin{tabular}{@{} r rrr rrr rrr rrr @{}}
    \toprule
    $\rho_S$ & \multicolumn{3}{c}{Raw} & \multicolumn{3}{c}{CUPED} & \multicolumn{3}{c}{CUPAC} & \multicolumn{3}{c}{DML DR} \\
    \cmidrule(lr){2-4} \cmidrule(lr){5-7} \cmidrule(lr){8-10} \cmidrule(lr){11-13}
     & Bias & SE r. & Cov. & Bias & SE r. & Cov. & Bias & SE r. & Cov. & Bias & SE r. & Cov. \\
    \midrule
    0.0 & $-0.9$  & 1.00 & 0.95 & $-0.9$  & 0.89 & 0.96 & $-0.4$  & 0.50 & 0.96 & $-0.3$  & 0.46 & 0.96 \\
    0.1 & $-2.3$  & 1.00 & 0.96 & $-2.4$  & 0.89 & 0.95 & $-1.9$  & 0.50 & 0.96 & $-1.8$  & 0.46 & 0.95 \\
    0.3 & $-5.3$  & 1.00 & 0.95 & $-5.4$  & 0.89 & 0.95 & $-4.9$  & 0.50 & 0.93 & $-4.8$  & 0.46 & 0.93 \\
    0.5 & $-8.3$  & 1.00 & 0.94 & $-8.4$  & 0.89 & 0.95 & $-7.9$  & 0.50 & 0.90 & $-7.8$  & 0.46 & 0.89 \\
    \bottomrule
\end{tabular}}

\vspace{0.5ex}
{\small \textit{Note}: SE r.\ = SE ratio relative to Raw; Cov.\ = coverage of the 95\% confidence interval for $\tau = 20$. 200 replications per regime.}
\end{table}

\end{document}